\documentclass[showpacs,preprintnumbers,amsmath,amssymb,floatfix,nofootinbib]{revtex4}
\usepackage[usenames]{color}
\usepackage{epsfig}
\usepackage{amssymb}

\newcommand{\be}{\begin{equation}}
\newcommand{\ee}{\end{equation}}
\newcommand{\bee}{\begin{eqnarray}}
\newcommand{\eee}{\end{eqnarray}}

\usepackage{ulem}
\def\dashuline{\bgroup
\ifdim\ULdepth=\maxdimen  
\settodepth\ULdepth{(j}\advance\ULdepth.4pt\fi
\markoverwith{\kern.15em
\vtop{\kern\ULdepth \hrule width .3em}
\kern.15em}\ULon}

\definecolor{grey}{rgb}{0.9,0.9,0.9}
\definecolor{black}{rgb}{0,0,0}

\def \irbaddress{Rudjer Bo\v{s}kovi\'{c} Institute, Bijeni\v{c}ka cesta 54, P.O. Box 180, 10002 Zagreb, Croatia}

\begin{document}

\title{ Reviving old, almost lost knowledge on T and K matrix poles and a link to the contemporary QCD spectrum }

\author{ A. \v{S}varc*}
\affiliation{\irbaddress \\
*E-mail: alfred.svarc@irb.hr}

\begin{abstract}
The old knowledge about interrelation among T-matrix, K-matrix and bare poles is summarized and put into modern perspective. 
\end{abstract} 

\maketitle

\section{Introduction} 
 The contemporary  QCD calculations of the baryon excitation spectra, defining resonances either as a proper value of the QCD Hamiltonian or as singularities of the QCD resolvent 
hadron Green function \cite{Pauli}, have in Aoki et al. \cite{Aoki2008} finally approached the physical pion mass limit with \mbox{$m_\pi = 156$~MeV}, but the comparison with the 
"experimental" nucleon mass spectrum is still done by comparing the lattice QCD resonant states with {\it Breit-Wigner parameters} \cite{ECT*}.  As summarized in the Prof. 
Hoehler's commentary in PDG1998 \cite{PDG1998}, the prevealing consensus has been reached that Breit-Wigner parameters are inherently model dependent. On the other hand, it  has  
for quite some time been well know in scattering theory \cite{Jang1995}, that Breit-Wigner parameters are nothing but an approximate parameterization of a scattering matrix pole.   
Therefore, now when the details of the spectrum start to matter, we claim that the comparison of lattice QCD results should be done not with Breit-Wigner parameters, but directly 
with much less model dependent set of resonance-quantifying parameters -- pole parameters. As scattering matrix poles are quantities lying in the complex energy plane, mass and 
half-width of a resonance are to be identified with the real and imaginary part of the pole position respectively,  and branching ratios are to be identified with the residua.   

The afore used definition of a resonance in QCD as a proper value of an \textit{in principle} non-hermitian Hamiltonian (hermitian Hamiltonians have real proper values, and poles 
are complex numbers) \textit{requires} that a QCD resonance is to be compared with a scattering parameter which corresponds \textit{to the same definition} of a resonance in the 
scattering theory. We have to be sure that we are talking about the same phenomenon at all.  I am not aware that the exact identification between QCD and scattering theory 
quantities can be done at all, but the closest we can get is to compare the singularities of the QCD resolvent hadron Green function  with the singularities of the scattering 
theory S-matrix. It is plausibly acceptable, but the exact proof is still missing. 

Now it seems like a very good moment to reopen an old question: " What a resonance actually is in the scattering theory, and what is its unambiguous signal?" 

A very intuitive definition, given in \cite{Moiseyev1998} and originating from Taylor textbook  \cite{Taylor}, says: \\
\textit{"Resonances are associated with metastable states of a system which has sufficient energy to break up into two or more subsystems."} 

However, this definition meets significant problems when has to be operationalized. A very natural connection between metastable states and a time delay in the scattering process 
has been offered and  discussed in the textbook by Bransden and Moorhouse \cite{BrandsdeenMoorhouse}, and there it has been particularly stressed that the existence of a time 
delay is only \textit{a necessary} condition for the existence of a resonance, and not a \textit{\textbf{necessary and sufficient}} one at the same time. This only means that two 
particles can show a significant time delay \textit{without} forming a resonance. Therefore, additional condition has to be imposed to eliminate all such surplus states. In 
Bransden and Moorhouse a resonance state is, therefore, associated with a situation when \textit{"...both the conditions, that of a time delay and that of a correspondence with a 
pole, are satisfied ..." }. Consequently, all conclusions deriving from only a time delay requirement (speed plot, quick backwards looping of the Argand diagram,...), are an 
indication that a resonance might exist, but are not a definite proof. Positive signals in these criteria mean that we may have a resonance, but to be sure that we do have, we 
have to eliminate all non-resonant states which produce similar effects. 

The strong step forward has, however,  been done in Dalitz and Moorhouse "What  is Resonance?" article \cite{DalitzMoorhouse} where it has been explicitly stated: \\
\textit{"...resonance phenomena are associated with the existence of eigenstates of the complete Hamiltonian for which there are only asymptotically outgoing waves."} 
\\  \indent
All other possibilities have also been analyzed in this paper, and the advantage of this definition has been stressed. This is, however, a definition which is in a perfect 
agreement with the present QCD understanding of resonances, so we strongly recommend it to the reader as an operationalization of the general definition. \\

Consequently we  affirm that in relating QCD with scattering theory parameters,  QCD resonant states should in principle be identified with scattering matrix poles. And now, this 
seems to be a good place to remind the reader that the same experimental data can be described by poles of several different functions: T-matrix poles \cite{Cut79,Zagreb,Arn06}, 
K-matrix poles \cite{Anisovicha,Anisovichb,Anisovich1} and  sometimes bare poles  \cite{Capstick2008}.  The aim of this article is to see their interconnection, and to infer which 
poles are to be linked directly to the QCD calculations, and which involve notable level of model dependence giving them limited physical significance. We believe that all poles 
can be identified with some QCD quantities at certain level of calculational approximations, but we shall try to show that the T-matrix poles are the ones which are "the optimal 
choice". Both, bare poles and K-matrix poles can be identified with some, let us call them "internal" resonant states, but these resonant states are not directly related to 
experimental data. In order to become observable they have to interact and get dressed. And this dressing procedure is essentially model dependent. We show that certainly all but 
T-matrix poles can be linked to QCD quantities only in a model dependent way, while model dependence for T-matrix poles is minimal. 

The immediate question arises: What are these "internal" resonant states? At this instant let me introduce them as a mere mathematical concept, a simplification of lattice QCD to 
a few body problem. Namely, we have two alternative approaches in QCD calculations: QCD models and lattice QCD. Lattice QCD is in principle many body theory, which assumes that 
the quarks, positioned on the corners of a grid interact via gluons exchanged along the grid lines, and the grid size is reduced. Such a theory does not know and does not care if 
quarks tend to make some kind of clusters on a microscopic level before becoming observable or not, it merely recognizes their final, macroscopic manifestation. On the other hand 
one may proceed through a model where we assume that quarks, before forming a macroscopic, observable state, form invisible, intermediate states which have to interact before 
becoming realistic. There is no reason why they should exist at all, but the analogy with other areas of physics  makes it plausible. Anyhow, scattering theory with T, K and bare 
matrices for which we shall show that each of them describes resonant structures at a different level, offers the tools to make a correspondence to a concept of QCD internal 
clustering.   

One should realize that there exists a  significant amount of old, but almost lost knowledge on definition of resonant signals and  scattering matrix singularity structure, and we 
feel that we are bound to refresh this knowledge prior to introducing new moments like immeasurability of off-shell effects due S-matrix invariance to filed transformations 
\cite{Fearing2000}, the effects which shed new light onto understanding the physical meaning of different type of scattering matrix poles.

The strategy of this article is to revive the old, almost lost knowledge on K and T matrix and bare poles, to elucidate their interplay and interpretation as resonant 
quantification signals, and put them into the modern perspective.
We shall show that generalization of the old K and T matrix ideas with the new field-theory concepts introduces a third set of pole parameters -- "bare poles", so the problem of a 
comparison  with QCD and lattice QCD gets even more versatile. 
We shall show that K and T matrix poles are closely, but {\it not straightforwardly} interconnected, and that their interconnection start seriously to depend on the manner how the 
resonant-background separation is implemented.  Namely, we shall demonstrate that the K-matrix poles within a certain approximation scheme may serve to quantify internal degrees 
of freedom, while the T-matrix poles should serve to describe external ones. We shall also show that the role of describing internal degrees of freedom by a third set of poles -- 
"bare poles" in the field theory sense, is extremely model dependent.  

 This is a convenient time to warn the reader about a very specific definition of bare states as  K-matrix poles introduced by Anisovich et al in mid 90-es 
\cite{Anisovicha,Anisovichb}. They have \textbf{\textit{defined}} bare states as K - matrix poles, being aware that such a definition includes a mass shift  due to all virtual 
$q\bar{q}$ interactions, and that this is different from the standard field theory definition where this mass shift \textbf{\textit{is not}} included. And the standard field 
theory bare mass is in their case simply called \textbf{\textit{"propagator pole mass"}}. In these first two papers this difference is explicitly discussed, and the reasons are 
explicitly given why the mass-shift corrections of the propagator pole mass in their model \textbf{\textit{should preserve}} the quark structure, and the bare mass in their 
definition should reflect the quark-level features. However, this discussion is almost entirely omitted in their successive papers \cite{Anisovich1} leaving the reader under the 
impression that K-matrix poles indeed are the bare masses in the field theory sense. Unfortunately, such a definition  has introduced a level of confusion into the physics 
community.   For decades  this difference is not fully recognized, so it is frequently met that the K matrix poles are directly identified with bare masses. However, it is an old 
knowledge that it is not so. 

At the same time we shall remind the reader of the fact that we should strongly distinguish between external resonant and external resonant-like behavior.  

Let us return to the main concept of the article. 
We shall discuss the interrelation of K and T matrix poles, and we shall show that it crucially depends on the type of the background contribution. Namely, for the constant 
background we shall show that there exists an 1-1 correspondence between T and K matrix poles, but as soon as the background becomes energy dependent (and what usually is the 
case), this convenient property is lost. One K matrix pole may induce more then one T - matrix poles which are all inexorably detected either as resonances, or as resonant like 
behavior of experimental observables. The only question is whether the examined poles will remain within the observable energy domain or not. However, applying the approximation 
that the energy dependent background may be represented as a meromorphic function of (at least in principle) infinite number of unphysical poles, we shall offer a simple mechanism 
how T-matrix poles may be analyzed, and how we may simply deduce whether a certain pole has a corresponding K matrix counterpart (genuine resonance) or is generated by an 
interplay of another  distant pole and energy dependent background (dynamic resonance).

At the same time, the article should serve as a stimulus to finally reorganize the PDG form in such a way to explicitly accentuate the model dependence of Breit-Wigner parameters. 
\section{Preliminary considerations}
Even before starting the discussion of scattering matrix poles as an authentic signal to be compared to QCD resonant structures, we shall make several introductory, preliminary 
considerations giving a precise definition of notions in scattering matrix formalism which are commonly taken as {\it well known}. 
\subsection{On resonances and resonant-like effects}
 First of all, we should be aware that a resonant-like behavior, i.e. the state of two or more particles which dwell in the vicinity of each other somewhat longer then necessary 
in a standard scattering process, can be achieved in ways which are not connected with the existence of resonance states. In such a case structures may be seen in differential and 
total cross sections, the scattering matrix {\it will show} necessary resonance conditions (prolonged time delay, peak in the speed plot), but the scattering matrix \textit{will 
not have a pole}. It is our task, and the intention of this paper, to strictly define which operators define resonances, and which define resonances {\it and} resonant-like 
effects.
\subsection{On genuine and dynamic resonances}
Not a small number of theorists believe  that there are two kind of resonances: genuine and dynamic. Genuine resonances are those resonant states generated by the  scattering 
matrix poles which are created by a pole of a more elementary internal resonant state function. Dynamic resonances are those resonant states generated by the  scattering matrix 
poles created dynamically, either as an interference of a distant genuine pole and a smooth, energy dependent background, or as an interference of two or more distant internal 
state poles. However, as a concept of \textit{a priori unmeasurable},   and hence model dependent internal states is in principle vague, it is not clear whether this separation is 
a mere mathematical convenience or a genuine physical fact.
\subsection{On the interrelation of K matrix with experiment} 
The question arises whether the K matrix is a direct representation of experimental data, or a derived quantity. 

In the reaction theory, experimental observables are directly given with the S-matrix matrix elements between initial and final state functions, with the elastic scattering 
interaction removed: $\langle f | S | i \rangle - \langle f | 1 | i \rangle$. And this quantity is directly proportional to the T-matrix. Consequently, single channel T-matrix 
matrix elements can be directly obtained by measuring one channel processes only. However, as the K-matrix is given as $K=i(1-S)/(1+S)$, to obtain the K-matrix matrix element 
$\langle f | K | i \rangle$ from experimentally extractable  S-matrix matrix element $\langle f | S | i \rangle$ one has to solve the equation $K=i(1-S)/(1+S)$, and that means 
inverting the S-matrix. Now we have to remember that the only correct way to treat scattering problem is a coupled-channel formalism, so S matrix  definitely has to be a matrix in 
the multi-channel space. Consequently, contrary to the T-matrix where for obtaining one channel matrix elements it is sufficient to measure only this particular channel, to obtain 
the K-matrix matrix element one has to invert the multi-channel S-matrix, and for that having a knowledge of a single channel is definitely insufficient. So, the K matrix is not 
directly measurable, it is a derived quantity.
\\ \noindent
{\it Let us summarize:} \\ 
The equations we are considering  are {\it matrix equations in the channel space}, so obtaining the single channel K matrix requires inverting the full multi-channel S(T)-matrix. 
And that of course means having the simultaneous  knowledge of all other channels. On the other hand, single channel T matrix can be obtained directly from experiment. 
So, in spite of seemingly legitimate parameterization of the experimental data by the K matrix form, such a parametrization has severe drawbacks in spite of being manifestly 
unitary. If the T matrix is conventionally taken to be a direct representation of experimental data (partial wave data), K matrix is obtained by the inversion of the 
\textbf{\textit{full}} multichannel T matrix, a matrix which is in principle poorly determined due to high level of ignorance about most but only few measured channels. 

The multi-channel K matrix tends to be instable. 
 This problem has been mentioned by several authors, but most clearly formulated by Cutkosky et al in ref. \cite{Cut79} where he suggested to use the K matrix method as one of the 
ways of parameterizing resonance parameters, but dropped it altogether because of high instability. A similar instability of multichannel fit where only a small subset of channels 
is known has also been discussed by Zagreb group, and most directly used when it has been shown that $\pi N \rightarrow \eta N$ channel data ensure the existence of P$_{11}$(1710) 
resonance \cite{CeciPRL06}.
\subsection{Pole and non-pole versus resonant and background separation}
The only {\it unique} way to single out the pole contribution from "everything else" is to make the Laurent expansion of the analytic function in the vicinity of the pole. Then we 
obtain energy independent pole parameters (pole position, residue) and energy dependent non-pole parts. Let us observe that in such a separation two very different type of 
contributions are included in the non-pole part. We have i) contribution of all other poles, and ii) contribution from all cuts and branch points due to the opening of inelastic 
channels. 

However, completely another story is if we want attribute some physical interpretation to these up to now only mathematical quantities .

For the K matrix it is straightforward. Namely, Laurent expansion of the K matrix is given as: 
\begin{eqnarray}
K(z) = \underbrace{\frac{r_k}{z-z_k}}_{K_p(z)} \  + \ K_{np}(z)
\end{eqnarray}
where $z_k$ is the K-matrix pole and $r_k$ K matrix residuum. If $K_{np}(z)$ is real on the physical axes, unitarity is manifestly maintained, and the pole and non-pole part can 
be given a reasonable physical interpretation as being resonant and non-resonant part: 
\begin{eqnarray}
K_p (z)=K_{Res} (z) \ ; \ \ \ K_{np}(z)=K_{bg}(z).
\end{eqnarray}
\noindent
However, for the T matrix the reasoning is somewhat more complicated.
\\ \\
Namely, if we make a Laurent expansion for the T matrix:
\begin{eqnarray}
T(z) = \underbrace{\frac{r_t}{z-z_t}}_{T_p(z)}\  + \ T_{np}(z)
\end{eqnarray}
then the unitarity, which we must require if we want to attribute some physical meaning to $T_p (z)$ and $T_{np} (z)$, is not manifestly maintained. One of many ways how to make a 
physically sensible separation in resonant and non-resonant part, is to replace the Laurent expansion with a manifestly unitary form, identical to the $T_p (z)$ and $T_{np} (z)$ 
in $z=z_t$, but modified elsewhere. We choose:
\begin{eqnarray}
T(z) = \underbrace{\frac{ {\rm Im}\  z_t}{z-z_t}}_{T_{Res}(z)} S_{bg}(z)+ T_{bg}(z)
\end{eqnarray}
Let us observe that in such a manner, the energy independent residuum of the Laurent expansion part $T_p (z)$ is effectively transformed to energy dependent function $ [{\rm Im}\  
z_t \cdot S_{bg}(z)$], which is traditionally called energy dependent partial width. 
\\ \\
Both forms of the singularity separation are used for different tasks. 
\\ \\
Pole non-pole separation is used when the pole positions are looked for \cite{Ceci-regularizacija2008,Doering2009}, while the resonant-background separation is used when the 
experimental data are being fitted within a certain model in order to obtained energy dependent partial waves. 

\section{Old, almost lost knowledge on the interplay between K and T matrix poles}  

We nowadays  have the whole plethora of dissonant attitudes towards the use of K matrix poles as a link between scattering and QCD resonances.  There is a variety of standings  
ranging from their complete and unequivocal  acceptance as bare masses in meson physics \cite{Anisovicha,Anisovichb,Anisovich1}  (however keeping in mind the very specific bare 
mass definition), to their total abolishing in some of the baryon states considerations \cite{Arndt09}. It is the ideal time to revive some almost lost knowledge about the 
interplay of K and T matrix poles as resonant signals in Breit-Wigner type approaches and fully define the correct use of K-matrix poles as a representation of bare masses. 

The first out of several non-trivial controversies is the interconnection of T and K matrix poles. We shall show, at least in a model almost maximally general,  that a one-to-one 
correspondence between the two indeed does exist, but the K-matrix pole position is {\it strongly} dependent on the T-matrix background. We find that the K-matrix pole can either 
be "in the vicinity" of the T-matrix pole  (in the observable physical region), and we shall call this pole genuine, or as for instance in a case of almost completely imaginary 
background, it will be shifted far away into the unphysical region, and the pole will be dynamic (generated by the distant or background effects).

However, we shall show that this convenient one-to one correspondence gets ultimately violated for the most general form of the energy dependent background.

In addition, we shall also demonstrate that the T matrix poles in general describe the external degrees of freedom, or in the terminology introduced afore describe all resonant  
effects. On the other hand we shall show that K matrix poles, or more precisely bare poles as their generalization when (if) mass-shift corrections can be eliminated, describe 
internal degrees of freedom entirely.

\subsection{T and K matrix pole interrelation for a multi-resonance, no-background scenario in a one-level Breit-Wigner model}

As it has already been shown by. I. J. R. Aitchison in 1972. \cite{Aitchison72}, in a simplified, but still very realistic model T and K matrix poles are directly related, and 
both of them can be given a concise physical interpretation.

In the most general case, one can write the T-matrix element $T_{ij}$ for a transition between continuum states $i$ and $j$ via the overlapping resonances forming the intermediate 
states as:

\begin{eqnarray} 
  T_{ij}= f_{i\bar{\alpha}}G^{'}_{\bar{\alpha}\bar{\beta}}f_{\bar{\beta j}}.
\end{eqnarray} 
The elements of the propagator matrix $G^{'}_{\bar{\alpha}\bar{\beta}}$ are in given by:
\begin{eqnarray} 
G^{'}_{\bar{\alpha}\bar{\beta}}= (M_0-  V - w)^{-1}_{\bar{\alpha}\bar{\beta}},
\label{eq:Eq0}
\end{eqnarray}

where  $\omega$ is the energy,
 $M_0$ is a diagonal bare mass matrix,  $V$ is the most general  non-Hermitian interaction operator and $\bar{\alpha}$ and $\bar{\beta}$ are the bare states, states which 
characterize the resonance \textbf{\textit{before any coupling is turned on}} (either resonance-channel, or the direct channel states coupling).

As general mathematical theorem states that any complex matrix can be written as a sum of Hermitian, and skew-Hermitian matrix, we decompose the interaction operator:

\begin{eqnarray} 
V = V^{Herm} + V^{SkewHerm}
\end{eqnarray}

Now  if we define:
\begin{eqnarray}
M^{Herm}=M_0-V^{Herm} \\
V^{SkewHerm}= \ \imath \ \frac{\Gamma}{2}^{Herm}
\end{eqnarray}
we may regroup the propagator matrix element:
\begin{eqnarray} 
G^{'}_{\bar{\alpha}\bar{\beta}}= (M^{Herm} -  \ \imath \ \frac{\Gamma}{2}^{Herm} - w)^{-1}_{\bar{\alpha}\bar{\beta}}.
\label{eq:Eq1}
\end{eqnarray} 
By skipping the suffix "Hermitian" we obtain the starting equation from Aitchison \cite{Aitchison72}:

\begin{eqnarray} 
G^{'}_{\bar{\alpha}\bar{\beta}}= (M- \ i \ \frac{\Gamma }{2} - w)^{-1}_{\bar{\alpha}\bar{\beta}},
\label{eq:Eq1}
\end{eqnarray} 
where $M=[[m_{\bar{\alpha}\bar{\beta}}]]$ is a Hermitian mass matrix, and $\Gamma=[[\sum_{i} 2 \pi \rho_{i} f_{\bar{\alpha} i}f_{i\bar{\beta}}]] $ is a Hermitian width matrix. 
$\rho_{i}$ is the phase-space factor for the channel $i$, and the \mbox{$f_{\bar{\alpha} i}=\langle \bar{\alpha}| V_c|i\rangle$} are the channel-bare resonance matrix 
 elements; the sum $\Gamma$ is subject to the energy-conserving condition $E_i=\omega$. Because of hermiticity of the transition potential $V_c$, $f_{\bar{\alpha} i}$ turn out to 
be real. 
  
It should be noted that, contrary to what is implied in the last sentence of \cite{Aitchison72}, the entries 
in the mass-matrix $M$ are not simply the bare mass eigenvalues $m_{\bar{\alpha}}$, but include 
also the mass-shifts (both diagonal and non-diagonal) induced by continuum coupling, as calculated in the 
$\bar{\alpha}$ basis. That is, 

\begin {equation} 
M_{{\bar{\alpha}}{\bar{\beta}}}=m_{{\bar{\alpha}}}\delta_{{\bar{\alpha}}{\bar{\beta}}} - 
\Delta_{{\bar{\alpha}}{\bar{\beta}}}
\end{equation}

where 
\begin{equation} 
\Delta_{{\bar{\alpha}}{\bar{\beta}}}= {\rm{P}} \Sigma_i \int \frac{\rho_i f_{{\bar{\alpha}}i} 
f_{i{\bar{\beta}}}}{E_i - \omega} dE_i \ .
\label{eq:new}
\end{equation} 

P stands for principal value. For more details see old papers by Feshbach, Kabir and  Stodolsky \cite{Ian1}
 From these formulae it is clear that $\Delta$ represents the mass-shift 
associated with virtual  (energy non-conserving) transitions to the continuum. 

The hermiticity of the mass itself also reflects a fact that the mass matrix $M$ is not a bare mass matrix, but contains the virtual state mass shift. Namely, if the mass matrix 
$M=[[m_{\bar{\alpha}\bar{\beta}}]]$  were the "bare mass matrix", it would be diagonal in the $\bar{\alpha}$ basis. But it certainly is not! It's a general Hermitian matrix. The 
matrix $M=[[m_{\bar{\alpha}\bar{\beta}}]]$  is the matrix whose entries are: (a) the bare mass eigenvalues; and (b) all the mass-shifts (diagonal and off-diagonal) induced by 
continuum coupling, as calculated in the $\bar{\alpha}$ basis. It is true that at the very end of the article this difference is not sufficiently stressed, but the whole 
derivation makes it straightforward that the mass matrix $M$ contains mass-shifts already. 
   
Aitchison has assumed a physical situation in which there are several overlapping resonance states with the same spin-parity and other relevant quantum numbers labeled A,B,C,... 
coupling to various two-particle continuum states labeled i, j, k... No background contributions are assumed at this level. 

From the very beginning, it is very important to make it crystal clear that we shall in principle have three different sorts of base states (hence three different pole 
parameters): \\ \indent a) bare states \{$\bar{\alpha}, \bar{\beta}....$\}, states which characterize the resonance before any coupling is turned on (either resonance-channel, or 
the direct channel states coupling); \\ \indent b) diagonal mass matrix states \{$\alpha, \beta,...$\}, states which are diagonalising the Hermitian mass matrix $M$, and \\ 
\indent c) the physical states \{$A,B,...$\}, states which are diagonalising the full T-matrix $M - \ i \frac{\Gamma}{2}$. \\ Observe, that Mass matrix $M$ and the Hermitian width 
matrix $\Gamma$ do not necessarily commute, so they can not be simultaneously diagonalised, but their sum, of course, can.  Consequently, singularities characterizing each set of 
the three sets of states will be different.   
\\ \\
Let us show their interrelation.
\\ \\
Mass matrix $M=[[m_{\bar{\alpha}\bar{\beta}}]]$ is a Hermitian operator, so it can be diagonalized in a new basis obtained by an unitary transformation $U$:
\begin{eqnarray} 
U_{\alpha \bar{\alpha}}M_{\bar{\alpha}\bar{\beta}}U_{\alpha \bar{\alpha}}^{-1} = m_{\alpha} \delta _{\alpha \beta}, \ \ \   UU^{\dagger}=I 
\end{eqnarray}
and new channel-bare resonance couplings are:
\begin{eqnarray}
f_{\beta j}=U_{\beta \bar{\beta}} \ f_{\bar{\beta} j}, \ \ \ f_{i \alpha}= f_{i \bar{\alpha}} \ U_{\alpha \bar{\alpha}}^{-1}.
\end{eqnarray}
Then the T-matrix in the new basis has the form.
\begin{eqnarray}
 T_{ij}& =& (f_{i \bar{\alpha}} \ U_{\alpha \bar{\alpha}}^{-1}) \ (U_{\alpha \bar{\alpha}} G^{'}_{\bar{\alpha} \bar{\beta}} U_{\beta \bar{\beta}}^{-1}) \ (U_{\beta \bar{\beta}} \ 
f_{\beta j})  \\
     & = & f_{i \alpha} G^{'}_{\alpha \beta} f_{\beta j}.
 \end{eqnarray}
 where
 \begin{eqnarray}
  G^{'-1}_{\alpha \beta} & = & (U M U^{-1} - \frac{i}{2} \ U \Gamma U^{-1} -w) = G^{-1}_{\alpha \beta} - \Sigma _{\alpha \beta}, \nonumber \\
   G^{-1}_{\alpha \beta} & = & (m_\alpha - w)\ \delta _{\alpha \beta} \nonumber \\
   \Sigma _{\alpha \beta} & = & \frac{i}{2}\sum_{i} 2 \pi \rho_{i} f_{\alpha i}f_{i \beta}.
 \label{eq:Mdiag}
\end{eqnarray}  
  And solving this leads us directly to the Dyson-Schwinger equation of the general form:
  \begin{eqnarray}
  G^{'} & = & G + G^{'} \Sigma   G\nonumber \\ 
  G^{'} & = &  G + G \Sigma  G + G \Sigma G  \Sigma  G + \cdots 
  \end{eqnarray}
  For the T-matrix we finally obtain:
  \begin{eqnarray}
    T_{ij} & = & f_{i \alpha} G^{'}_{\alpha \beta} \ f_{\beta j}   \nonumber  \\
	  =    & &  f_{i \alpha} G_{\alpha \beta} \  f_{\beta j} + f_{i \alpha} \ G_{\alpha \gamma}  \Sigma _{\gamma \delta}  G_{\delta \beta} \ f_{\beta j} + \cdots\nonumber \\ 
      = & &  f_{i \alpha} G_{\alpha \beta} \  f_{\beta j} +  f_{i \alpha} \ G_{\alpha \gamma}\left(\sum_{k} i \pi  \ \rho_{k} f_{\gamma k}  f_{k \delta}\right) G_{\delta \beta} \ 
f_{\beta j}  \nonumber \\ & &  + \cdots \nonumber \\
  = & &  f_{i \alpha} G_{\alpha \beta} \  f_{\beta j} + i \pi \  \sum_{k} \ (f_{i \alpha}  G_{\alpha \gamma} f_{\gamma k}) \ \rho_{k} \ (f_{k \delta} G_{\delta \beta} f_{\beta j} 
)  
  \nonumber \\ & & + \cdots \nonumber 
  \end{eqnarray}  
Recalling that the sum over repeated indices is assumed we define the new function $K_{ij}$ as:
\begin{eqnarray}
  f_{i \alpha}  G_{\alpha \beta} f_{\beta j} & \stackrel{def}{=}&  K_{ij} 
 \end{eqnarray} 
 and we finally obtain:
 \begin{eqnarray}
 T_{ij} & =& K_{ij} + i \pi \ \sum_{k}  K_{ik} \ \rho _{k} \ K_{kj} + \cdots.
 \end{eqnarray}
 The only thing left is to recognize that this is the iterated form of a K-matrix definition:
 \begin{eqnarray}
 T=K(1-i\pi \rho K)^{-1}
 \end{eqnarray}
 Consequently, substituting $G_{\alpha \beta}$ from Eg. (\ref{eq:Mdiag})  we directly obtain:
 \begin{eqnarray}
 K_{ij} & =& \sum_{\alpha, \beta} f_{i \alpha}  G_{\alpha \beta} f_{\beta j}  =  \sum_{\alpha, \beta} f_{i \alpha} \frac{\delta _{\alpha \beta}}{m_\alpha - w} f_{\beta j} 
\nonumber \\
 & = & \sum_{\alpha} f_{i \alpha} \frac{1}{m_\alpha - w} f_{\alpha j} \nonumber.
 \end{eqnarray} 
 And that finally gives us the interrelation between eigenvalues of the Hermitian mass matrix M, and the K-matrix pole positions. 
 \\ \\
 They are namely equal.  
 \\ \\
 \textit{ \underline{Summary}: }
 
\textit{{\small 
The complete T-matrix, hence all of its poles, can be represented by a K-matrix represented as a sum of poles. 
 In this approximation (Hermitian width matrix in the Breit-Wigner formula, no background), K-matrix poles are identically equal to the mass eigenvalues of the Hermitian mass 
matrix and  give a direct insight into the internal resonant state distribution.} }
\subsection{Uniqueness of the interrelation of K and T matrix poles}
Up to now we have shown that the existence of K-matrix pole requires the existence of  T matrix ones. However, vice versa is not at all clear. 

In paper by L. Rosenfeld \cite{Rosenfeld70} it has been generally shown that in a constant background, multi-resonance case there is a 1-1 correspondence between K and T matrix 
poles.
As the intention of this paper is not to repeat all the mathematical intricacies of the full derivations, but to give a summary of procedures and a reference to the original 
paper, we shall just skim through the original derivation to give the reader the impression of the level of mathematical rigor. 
\\ \\
He starts with the multi-resonance, no-background form of the multi-channel S matrix:
\begin{eqnarray}
\label{eq:Smatrix}
S_{k'k} & = & \delta _{k'k} - i \sum _{n} \frac{u_{k'n}u_{nk}}{E- {\cal E} _{n}},
\end{eqnarray}
where $u_{kn}$ is a "complex partial width", and ${\cal E}_{n}$ is T-matrix pole.
\\ \\
In ref.~\cite{Rosenfeld70} it is shown that it is a matter of straightforward algebraic manipulation to obtain that indeed, K matrix defined by \mbox{$S=\frac{\delta \ + \ i \ 
K}{\delta \  - \ i \ K}$} can be represented in the form:
\begin{eqnarray}
K_{kk'} & = & - \frac{1}{2} \sum _{\mu} \frac{v_{k \mu}v_{\mu k'}}{E-\epsilon _\mu}
\end{eqnarray}
where $\epsilon _\mu$ are K matrix pole parameters, and $v_{k \mu}$ is a set of real parameters analogous to the partial width parameters $u_{k \mu}$, and obtained from them by a 
complex-orthogonal transformation $O=||O_{n \mu}||$ with $O^{\dagger}O=\delta$, and which is diagonalizing a set of symmetrical matrices $\Delta$ and $\nabla$ appearing as an 
intermediate step in the derivation. In a matrix notation these interrelations can for the T $\rightarrow$K transformation be written as:
\begin{eqnarray}
{\cal E} + \frac{1}{2} \sum _{k} u^{k \dagger}u^{k} = O \epsilon O^{\dagger}, \nonumber \\
u^{k}=v^{k}O^{\dagger}, \  \ \ u^{k \dagger} = O v^{k \dagger},
\end{eqnarray}
so the K matrix poles can be obtained by finding the roots of the characteristic equation:
\begin{eqnarray}
{\rm det} \left[ (E- {\cal E}) \delta _{\mu \mu'} + \frac{1}{2} i \sum _{k} u_{\mu k} u_{\mu k'} \right]=0.
\end{eqnarray}
Conversely, these interrelations can for the K $\rightarrow$T transformation be written as:
\begin{eqnarray}
 \epsilon + \frac{1}{2} \sum _{k} v^{k \dagger}v^{k} = O {\cal E} O^{\dagger}, \nonumber \\
v^{k}=u^{k}O, \  \ \ v^{k \dagger} = O^{\dagger} u^{k \dagger},
\end{eqnarray}
so the T matrix poles can be obtained by finding the roots of the characteristic equation:
\begin{eqnarray}
{\rm det} \left[ (E- \epsilon) \delta _{\mu \mu'} + \frac{1}{2} i \sum _{k} v_{\mu k} v_{\mu k'} \right]=0.
\end{eqnarray}
And this gives a 1-1 correspondence for the multi-resonance, no-background case.
\\ \\
\textit{ \underline{Summary}: }

\textit{{\small We have shown that in no-background scenario there exists a 1-1 correspondence between T matrix poles describing external degrees of freedom and K matrix poles 
describing possible internal structures.} } \newpage
\subsection{Generalization to more realistic cases}
However, life is not simple. The afore used approximations have to be strengthen by allowing for the realistic background. Therefore, we have to generalize this analysis 
accordingly. 

\subsubsection{Non-vanishing background}
We discuss three levels of including background contributions into considerations:
\begin{description}
\item[\it i)] the constant background; 
\item[\it ii)] energy dependent background represented as a sum of finite number of "unphysical" poles, and 
\item[\it iii)] fully energy dependent background.
\end{description}  
  In each of the three cases, the background contribution influences the K matrix pole position, but in first two case the 1-1 correspondence between K and T- matrix poles is 
retained. Presence of the background effectively does shift the K matrix pole position, and this shift can in principle be big, but for each T-matrix pole we do have a 
corresponding K matrix pole, regardless of the fact that due to the presence of the background it may be shifted far into unobserved energy domain. 
However, the size of this "background" shift allows us to clearly distinguish between two type of T-matrix poles: ones for which we shall find corresponding, nearby K-matrix 
poles, and ones for which such a pole could not be clearly identified. The first type of T-matrix poles we call genuine resonances, while the latter ones are dynamic ones, ones 
which do not have a related bound state singularity, hence ones which describe resonant-like behavior only. 

We shall finally discuss the third possibility, the possibility which in principle allows that 1-1 correspondence is spoiled, and show that the main conclusions about K-T poles 
interrelation hold regardless of the possible realization of this most general case. 
 \\ \\ 
{\it \underline{i. Constant background}} \\ \\
There is an old, but well documented procedure to treat the constant background. In old paper by K. M. McVoy \cite{McVoy69} it has been shown that the scattering matrix unitary 
can help us to eliminate background terms using the background dependent unitary transformation, and end up with a no-background form of resonance representation of scattering 
matrix, but with the redefined resonance parameters. As K. M. McVoy has analyzed the 2-resonance many channel case only, we shall preferably present the more general case given by 
L. Rosenfeld one year later \cite{Rosenfeld70}.

Let us assume that the collision matrix between two channels $c'$ and $c$ is given by:
\begin{eqnarray}
U_{c'c}= B_{c'c} - i \sum _{n} \frac{g_{c'n}g_{nc}}{E- {\cal E}}
\label{eq:Umatrix}
\end{eqnarray}
instead with Eq.~(\ref{eq:Smatrix}), where $B_{c'c}$, $g_{c'n}$ and $g_{nc}$ are complex constants. Then the background parameters can be eliminated following K. M. McVoy 
\cite{McVoy69}. The unitarity of the matrix $U$ enforces that the matrix $B$ is unitary as well. Consequently, if we reduce the terms of the expression (\ref{eq:Umatrix}) to the 
same denominator, the matrix $B$ appears at the coefficient of the highest power E in the numerator. The eigenvalues of $B$ may therefore be written in the form $e^{2i \beta_k}$; 
since $B$ must also be symmetrical the corresponding eigenvectors $\chi _{ck}$ may be taken to be real, and if they are normalized to unity, they form an orthogonal matrix. The 
matrix b defined by:
\begin{eqnarray}
b_{ck} = e^{i \beta_{k}} \chi_{ck}
\end{eqnarray}
is then unitary, and it yields for B the expression $B=b b^\dagger$. We have now only to define the transforms $u_{kn}\equiv u_{nk}$ of the partial width parameters by the inverse 
matrix $b^{-1}$:
\begin{eqnarray}
u_{kn}\equiv u_{nk} = \sum _{c'} g_{nc'} b^*_{c'k},
\end{eqnarray}
in order to bring the matrix $U$ into the form 
\begin{eqnarray}
U= b \ S \ b^\dagger
\end{eqnarray}
end we explicitly obtain the equation (\ref{eq:Smatrix}).
\\ \\
Consequently, the form of the equation is maintained with the redefined coefficients, and the number of T and K matrix poles still uniquely correspond. However,  the K matrix pole 
position becomes background dependent.
\\ \\
{\it \underline{Digression:}   K and T matrix pole interrelation in SAID}
\\ \\
It seems like a perfect time to make a small, but important digression on the role of background parameterization in K-matrix pole identification. 
\\ \\
It may seem that previous considerations are in some form of disagreement with the SAID (GWU/VPI) partial wave analysis procedure.  Namely, on numerous occasions GWU/VPI group has 
claimed \cite{Arndt-polinomial background} that their formalism is able to generate T-matrix poles in spite of the fact that they parameterize the K matrix contribution only in a 
form of energy dependent polynomials without K-matrix poles whatsoever. All in all, they claim that they can generate well defined, strong T matrix poles without any K-matrix 
ones, and that would violate all our previous statements \footnote[1]{Since I have started to write this text a considerable modification of this statement has been done by one of 
the most recent acquisitions of GWU group  Mark Paris. He has pointed out that GWU formalism basically starts with no-pole Chew-Mandelstam K matrix representation, and that it can 
spontaneously generate  standard Heitler K-matrix poles \cite{Paris2010}. However, the aim of this exercise is to show that even no-pole  Heitler K-matrix assumptions can generate 
T-matrix poles in the physical region.}.
\\ \\
But we shall show that it is not so. They implicitly do have K-matrix poles, but due to the T matrix resonant-background interplay,  these poles are shifted \textbf{\textit{far 
away from}} the energy range of relevance, so their fits do not see their contribution. 
\\ \\
Let us give us a very simple example which will demonstrate what happens. 
\\ \\ 
Let us take that the K-matrix is only linear in energy:
\begin{eqnarray}
\label{eq:27}
K(w)  & = &  \alpha -w 
\end{eqnarray}
One can simply show that  such a K- matrix produces a perfectly legitimate T-matrix pole in $(\alpha + i)$, but in the presence of a purely imaginary background:
\begin{eqnarray}
\label{eq:28}
T(w)  & = & -\frac{1}{w-(\alpha + i)} \ + \ i 
\end{eqnarray}
One would at a first glance say that a T matrix pole does not have a corresponding K matrix one. However, we do have a K matrix pole, but due to the background-resonance interplay 
it  has been shifted into infinity. It returns from infinity immediately when the T matrix background acquires real part. 

To show this let us unitary rotate the fully imaginary background by the infinitesimal angle $\epsilon$. Remembering that unitary rotation of a T matrix is realized using the 
formula $T'(w)= T(w)S_B+T_B$, we obtain:
\begin{eqnarray}
\label{eq:29}
T'(w)=T(w)e^{  2 \, \epsilon \, i }+\frac{e^{  2 \, \epsilon i}-1}{2i}.
\end{eqnarray}
The corresponding K' matrix will have a very indicating form:
\begin{eqnarray}
\label{eq:30}
K'(w) &=& \frac{ (\alpha - w)\cos \epsilon + (\alpha - w ) \sin \epsilon}{ \cos \epsilon + (w-\alpha) \sin \epsilon}   \nonumber \\
& + & \ 0 * i. 
\end{eqnarray}
As expected, K matrix is a real function and  has a pole at $w=\alpha - \cot \epsilon/2$. Consequently, for $\epsilon=0$ pole is at -$\infty$, and when $\epsilon$ increase slowly 
approaches the physical domain.  
\\ \\
There is also another quick way to get Eq.~(\ref{eq:30}).  Namely, one recalls that formula Eq.~(\ref{eq:29}) is the usual expression for modifying a given $T(w)$ by the addition 
of a background phase $\epsilon$ .   Eq.~(\ref{eq:29})  is  saying that the total phase is just $\delta + \epsilon$, where $\delta$ is the phase shift in $T(w)$. Now according to 
Eqs.~(\ref{eq:27}) and (\ref{eq:28}), the K  matrix is just $\tan (\delta$). So K' is $\tan (\delta + \epsilon)$, which gives $K'=(K+ \tan \epsilon)/(1 - K \tan \epsilon) = 
(\alpha - w + \tan \epsilon)/(1-[\alpha - w] \tan \epsilon)$ immediately. The usual phase space factor  $\rho$ is because of simplicity included in the K matrix definition.
\\ \\
As a conclusion, when SAID starts with a polynomial expansion for the K matrix, they basically choose a very "close to imaginary" form of the background contribution, they create 
their poles dynamically, and never actually find the distant unphysical K matrix poles. \\ \\  
\underline{\it ii. energy dependent background represented as a sum of finite number of "unphysical" poles}
\\

Theoretically, background term is in principle energy dependent function of energy. However, an approximation is very often used that one may decompose the background term into a 
finite number of  unphysical pole terms, and proceed as for the multi-resonant case with constant background. This approach has been introduced and defended by Cutkosky et al 
\cite{Cut79} in CMB PWA and extensively by Zagreb group \cite{Zagreb}.

The idea is simple. An energy dependent, expectedly smooth background is represented as a finite sum of poles regardless of the fact that they bear absolutely no physical meaning 
whatsoever. That approach is very often looked at with non-hidden amount of distrust, but no one has been able to demonstrate any roughness of such an approximation. Therefore, we 
believe that even if not completely rigorous, this approximation will give a sensible idea about the character of each individual resonance. Namely, the position of poles 
mimicking background, and most of all their significance for the amount of shift of the K matrix pole with respect to the T matrix one, will characterize each resonance. 

Consequently, one-to-one correspondence between T and K matrix poles is maintained up to a certain level. In a world where all energy dependence of the non-pole contributions is 
mimicked with a number of unphysical poles, the one-to-one correspondence between T and K matrix poles is maintained, but this time {\it not for} one {\it but for two} type of 
poles: one describing the genuine singularities, and ones describing the background contribution. 

Let us elaborate on what we have actually done when we have assumed that the background {\it indeed is} a meromorphic function. In that case, instead of solving a nonlinear 
equation for the T-matrix pole positions when the background has a general energy dependent form, we have  actually introduced a new type of K-matrix poles into the "story", poles 
which "mimick" the background. And here we rely on the approximation that the solutions of a nonlinear equation with general type background will be close to the solutions of 
"pole-type" equation when the meromorphic function representing a general background is very close to the original function. And that is equivalent to saying that the K 
$\rightarrow$ T matrix transformation equation with an energy dependent background has more then one solution for only one K matrix pole (hardly a new fact for the well informed 
reader), but now we may understand the additional solutions as \textbf{\textit{new type}} of T-matrix poles - poles  \textit{originating from unphysical K matrix poles} introduced 
in order to describe the energy dependent background.  Now we do not have one, but two type of T-matrix poles "in the game": i) poles which correspond to the real internal 
singularities; and ii) poles which correspond to the background. We call the first poles genuine poles, and the second type we call dynamic ones.

So, instead of saying that for an energy dependent background the one-to-one correspondence between K and T matrix poles is lost, we may actually say that the one-to-one 
correspondence is maintained in a restricted sense: it is manifestly maintained for all T matrix poles, but for genuine ones the related K matrix pole is "nearby", while for the 
dynamic ones (like the ones generated by the linear K matrix) the corresponding pole is far away in unphysical range.  The real problem is, of course, that both type of poles are 
"seen" from the experimental side in exactly the same manner. So, in principle we have three "scenarios" in the game depending on how the dressing procedure shifts the  two type 
of internal poles : \\  i) genuine poles are shifted into the observed domain, while background poles are pushed far into the unphysical part; \\  ii) genuine poles and part of 
the background poles are both shifted into the observed domain; and \\ iii) genuine poles are shifted outside observed domain, while part or all poles remain within. 
\\ \\
The speculative question is: {\it "How realistic the meromorphic background approximation  actually is?"} 
\\ \\
The answer is non-trivial. It is correct up to a good approximation.
\\  \\
\underline{\it iii. general form of energy dependent background} \\ \\  \indent
Going on to the most general case, to the situation when the background is a nonmeromorphic function containing the genuine cut, is a speculative extrapolation of the meromorphic 
approximation, and is again based on old and not so generally known knowledge. 

To replace the background with a meromorphic function with a finite number of poles has been tried occasionally, but in ref \cite{Brune96} it has been clearly stated  that {\it 
"the background can not be adequately represented by sum of real-energy poles and a constant"}. However, an old paper by Kaufman yet from 1963 \cite{Kaufman63} clearly says that 
if not finite, then an infinite sum of poles suffices for a confident representation.  

\noindent
So, a conclusion emerges: \\ The only thing which is changed with respect to the meromorphic background case is  that the number of poles needed to represent the background 
increases.  Everything else remains the same.

\noindent
Let us summarize: \\
Introducing background contributions influences more intricate interrelation between K and T matrix poles. Constant and energy dependent background represented as a sum of finite 
number of poles maintain uniqueness, however two kinds of poles emerge: genuine and dynamic.  General form of energy dependent background offers multi-valued T matrix poles for a 
given K matrix one, but each of these poles lies on a different Riemann sheet. The obvious advantage of representing the background with a meromorphic function is that one retains 
a full control over the number of poles, or in other words, over the number  of solutions of a nonlinear equation an energy dependent background is imposing upon us. 
\\ \\
{\it Corrolary} \\ \\
Carnegie-Melon-Berkeley type models represent the background contribution in a multipole form \cite{Cut79,Zagreb,Vra00}, consequently they maintain one-to-one correspondence 
between K and T matrix poles.
Other coupled-channel models \cite{EBAC,Giessen,GWU-VPI,Bonn-Gatchina,Mainz-Dubna,Juelich} allow for a general energy dependence of the background contribution, so singe internal 
structure pole may produce multipole T matrix ones, indicating the corresponding resonant-like behavior. Such a resonant-like behavior in CMB type models is observed when the 
internal structure poles, corresponding to a chosen T matrix one lie far in the unphysical region of the complex energy plane. 

\section{On a link between scattering matrix poles and QCD}
Up to now we have fully explained the interconnection of bare, K and T matrix poles, and their interdependence regarding the number of poles and type of background contributions. 
We have established the fact that different scattering matrix poles (bare, K matrix and T matrix poles), even when connected, really describe different things.  So it is clear 
that they can only  be compared with different aspects of QCD calculations. As it has never been exactly stated which poles should be connected to which type of QCD calculations, 
let us give you our correlation scheme. 

 We claim that poles describing internal degrees of freedom should be compared with QCD models, and external ones should be compared to lattice QCD. Internal degrees of freedom 
are by definition model dependent, and only external ones can be obtained directly from experiment. 

A tentative scheme of QCD - scattering theory connection,  based on the level to which the quark loops are effectively included in the mass calculation, is offered: \\ \\
NO \vspace*{0.1cm} LOOPS: \\
\indent {${\rm QCD} ^{\rm constituent} _{\rm models} \ \Longleftrightarrow \ {\rm bare \ poles}$}   \\ \\
ALL VIRTUAL \vspace*{0.1cm} LOOPS: \\ 
\indent {${\rm  QCD} \  ^{\rm unquenched} _{\rm models} \   \Longleftrightarrow  \ {\rm K \ matrix \ poles}$}  \\ \\
ALL VIRTUAL AND REAL \vspace*{0.1cm} LOOPS: \\ 
\indent {${\rm QCD \ lattice }    \Longleftrightarrow \ {\rm T \ matrix \ poles}$}   \\

So, in principle, it would seem that the full answer is given. However, it is only theoretical.

It is not at all clear which quantities on QCD  and scattering theory sides can be extracted in a model independent way, and we have to  suggest a model independent point of 
contact between scattering theory and and QCD.  

From the scattering theory side, we know that Breit-Wigner parameters are extremely model dependent, so they are not even offered as an option. 

T-matrix poles at the present moment seem to be the scattering theory quantity with lowest level of model dependence involved. Some assumptions about the analytic form on the 
level of the input functions have to be made, but still it seems that  the overall agreement about pole positions is soon to be reached \cite{SvarcSeattle}. On the other hand, 
T-matrix pole positions are far from being reliably predicted from lattice QCD. 

The completely opposite side of the spectrum are bare parameters. Models on the QCD side can easily be constructed, not so easily solved, but still we do have numerous predictions 
of different constituent and quenched QCD models (see ref. \cite{Capstick2000}). However, extracting bare parameters from the experimental data involves identifying 
$q\bar{q}$-mass shifts, count it out, and extract bare masses. This procedure certainly involves model dependence as has been shown recently by Fearing and Scherer 
\cite{Fearing2000}. 

Standing somewhere in-between are the K-matrix poles. However, as discussed in II.C, K-matrix poles are also not a directly measurable quantity, so their values are much more 
difficult to be extracted from experimental data as one needs the knowledge of (theoretically) all channels to be able to invert the S-matrix. We admit that it is feasible, but 
multichannel K-matrix pole extraction is still limited to worryingly few number of channels. On the other hand it is quite complicated to introduce all possible $q\bar{q}$ 
corrections to the quenched QCD models. Of course, the only question which has to be addressed very carefully is: "To which level of precision are all off-shell meson 
(meson-baryon) corrections included into the models?"

In all three cases the situation is far from simple, so let us analyses what has been done in each of the sectors. 

\subsection{Internal structures}
Regarding internal degrees of freedom, a very important analysis has been made by Anisovich et al. in ref. \cite{Anisovichb}. In Eq. (2.28) of this reference it has been shown 
that the K-matrix can be separated in two additive parts: a) pole contribution whose mass $\mu$ is the bare mass $m^{bare}$ in the field theory sense corrected with the 
$q\bar{q}$-mas shift $\Delta$, and b) contribution of the physical meson-meson processes $f_{ab}$: \\ \centerline{\mbox{ $K_{ab}(w)=\frac{g_ag_b}{\mu-w}+f_{ab}; \ \ 
\mu=m^{bare}-\Delta$.}}

So, both: K matrix poles and bare poles do reveal the internal structure, but represented on a different level of calculation. On the other hand, going from the internal structure 
poles to the T-matrix poles (which indeed are the experimentally accessible quantity)  is a process completely determined by the real meson-meson contribution function $f_{ab}$ 
which surfaces as the K-matrix background. Hence, for the discussion of internal structure poles interrelation with the experimentally attainable T matrix ones,  all conclusions 
made for the K-T poles interrelation exposed in former chapters can be directly applied. 
\subsubsection{Bare parameters}
Quenched QCD models have been constructed for quite some time (see a review by Capstick et al. \cite{Capstick2000}), but in the absence of a better candidates, their predictions 
have always been compared to Breit-Wigner parameters as given by Particle Data Group \cite{PDG}. The danger of Breit-Wigner model dependence is known \cite{Capstick2008}, but not 
much has been done to overcome it. 

However, there is a number of approaches trying to correlate other, \textbf{\textit{non Breit-Wigner quantities}}, to the quenched QCD model predictions.  
A long time ago, a bare parameters from dynamical, coupled channel models based on the effective Lagrangian approach have been recommended as the direct signal of internal degrees 
of freedom (for the overview of models see review by Matsuyama, Sato and Lee \cite{MatSatLee07}), and the proposition has been made to interpret them as a quenched quark model 
resonant states \cite{SatoLee}. 

Following the similar logic, the attempt to use the bare parameters from the coupled-channel Carnegie-Melon-Berkeley model \cite{Cut79}, for the same purposes has been made by 
Zagreb group \cite{CeciEPJC09}.

Both groups have shown a significant level of resemblance between QCD quenched  model parameters on one side, and scattering theory bare parameters on the other. 

However, such an interpretation has raised quite some controversies because of unclear link between meson degrees of freedom in scattering theory and quark models because of the 
invariance of effective Lagrangian theories to the field transformations \cite{Hanhart}. 

 Right now it seems that separating bare mass and $(q\bar{q}, \ q\bar{q}q q q)$-mass shift corrections can not be done in a model independent way, because in effective Lagrangian 
approach all of-shell intermediate state corrections of one Lagrangian can be transformed into the coupling constant of another, point-like one \cite{Capstick2008}.

However, it really does not mean that there does not exists such a point like Lagrangian whose bare masses correspond to the K-matrix poles because its $(q\bar{q},\ q\bar{q}q q 
q)$ corrections are non-existent. However, as the separation of bare properties and dressing is basically arbitrary, \textbf{\textit{we repeat that it does not have any physical 
meaning whatsoever.} }.  
\subsubsection{K matrix poles}
K-matrix pole analysis has been extensively used by Anisovich et al \cite{Anisovicha, Anisovichb,Anisovich1}, and in meson physics in general \cite{Klempt}.  However, one has to 
bare in mind that a bare mass definition includes the  $q\bar{q}$ corrections, so a comparison to QCD models have been done under assumption that the mass shifts preserve the QCD 
structure. 
\\ \\ \noindent
\underline{\it Summary:} \\ 
 
Both, bare an K matrix poles reflect the internal pole distribution to a certain degree, but it should be bery clearly distinguished to which level are all intermediate-state 
off-shell corrections of  $(q\bar{q}, \ q\bar{q}q q q)$ type are included in certain quark model. Due to the problems of extracting bare and K-matrix poles from the experimental 
data, we strongly believe that non-trivial steps should be undertaken before they are to be connected to some specific constituent quark model. We refer the reader to a nice 
research project description about the presently available possibilities, given at 5th PWA Workshop in ECT*-Trento by  H. Haberzetl \cite{HaberzetlTrento}.

\subsection{External structures}
External structures are described exclusively by T-matrix poles. 
T matrix poles are extracted from the experimental quantities measured on the real, physical axes, and obtained analytic functions are extrapolated into the complex energy plane.  
If background is assumed to be either constant, or given in a multipole expansion form, the number of internal poles correspond to the number of external ones. However for the 
energy dependent background  one internal pole can produce a number of external ones, depending on the number of open channels. The multi-channel character of the problem is of 
\textbf{\textit{decisive importance}}, because the lack of channels introduces ambiguities in pole determination. Evenmore, in the case of insufficient number of channels, some 
poles may remain undetected.  

If we assume that an energy dependent background can be replaced by an (in)finite sum of poles, the number of T matrix poles again corresponds to the number of K matrix poles 
(bare poles), but  some of the internal poles \textbf{\textit{may be far away from the observable energy range.}} 

The lattice QCD calculates the mass spectrum of all possible many-quark interactions, so it is clear that it should be compared to the "external structures" -- T matrix poles. 
\section{Conclusions}
Summarizing and modernizing the existing knowledge on few-body-scattering singularities we conclude:

\begin{enumerate}
  \item There exists a 1-1 correspondence between K and T matrix poles for two types of background: energy independent, and energy dependent (meromorphic form). 
  \item   In the case of energy dependent background of general functional form, this 1-1 correspondence is lost.
  \item  If an infinite number of poles is introduced to represent the background, the 1-1 correspondence is restored. 
  \item Two type of T-matrix poles are recognized: 
   \begin{description}
          \item[i)] genuine (corresponding to the nearby bare/K matrix pole) 
		  \item[ii)] dynamic (when the nearby bare/K matrix pole can not be identified) 
    \end{description}
  \item T matrix poles describe both, resonant {\it and} resonant-like behavior, and as only quantities which are (in principle) directly accessible from experiment are to be 
introduced as a comparison point for QCD calculations. 
\item All scattering matrix poles can be related to QCD resonant structures at certain level;
  	\begin{description}
          \item[i)] bare and K matrix poles to constituent quark models with the approximation of infinitely long bound states
		  \begin{description}
                  \item[a)] bare poles to constituent models in which \textbf{\textit{NO}} intermediate state energy shift corrections are taken into account
		          \item[b)] K matrix poles to constituent models in which \textbf{\textit{ALL}} intermediate state energy shift corrections are taken into account
          \end{description}
		  \item[ii)] T-matrix poles to lattice QCD corrections where intermediate state decay is taken into account
    \end{description}

  \end{enumerate}

\section*{Acknowledgments} 
I am particularly grateful to Prof. Dr. Ian Aitchison for a detailed critical review of the manuscript in a certain stage which resulted in significant improvement of the clarity 
of the manuscript. I stress that he has helped me a lot to eliminate some misconceptions, and focus the article on elucidating some very important aspects of the problem. I would 
also like to add that this review has been written with the help of  MZO\v{S}-DAAD bilateral agreement, and the help of ECT* Trento Center is also recognized.  I am also very 
grateful to all other colleagues of mine who have discussed issues connected with this write-up, but I am  in particular thankful to Prof. Dr. Siegfried F. Krewald from Juelich 
who encouraged me to publish this manuscript even only in the preprint form.

\end{document}